\documentclass[aps,preprint,preprintnumbers,amsmath,amssymb,nofootinbib]{revtex4}
\usepackage{amssymb}
\usepackage{lscape}
\usepackage{amsmath}
\usepackage{bm}
\usepackage{graphicx}
\usepackage{epsfig}

\begin{document}
\title{Inflationary gravitational waves from unified spinor fields}
\author{$^{2}$ Luis Santiago Ridao\footnote{E-mail address: santiagoridao@hotmail.com}, $^{1,2}$ Marcos R. A. Arcod\'{\i}a\footnote{E-mail address: marcodia@mdp.edu.ar}, $^{2}$ Jes\'us Mart\'{\i}n Romero\footnote{E-mail address: jesusromero@conicet.gov.ar} and $^{1,2}$ Mauricio Bellini\footnote{E-mail address: mbellini@mdp.edu.ar}}
\address{$^1$ Departamento de F\'isica, Facultad de Ciencias Exactas y
Naturales, Universidad Nacional de Mar del Plata, Funes 3350, C.P.
7600, Mar del Plata, Argentina.\\
$^2$ Instituto de Investigaciones F\'{\i}sicas de Mar del Plata (IFIMAR), \\
Consejo Nacional de Investigaciones Cient\'ificas y T\'ecnicas
(CONICET), Mar del Plata, Argentina.}
\begin{abstract}
Recent observations of Gravitational Waves (GW) generated by black-hole collisions have opened a new window to explore the universe in diverse scales. Detection of primordial gravitational waves is expected to happen in the next years. However, the standard theory to describe these effects was developed for weak gravitational waves, when their dynamics can be linearized. In this work we develop a non-perturbative formalism to describe GW using the Unified Spinor Fields (USF) theory. The tensor index is calculated and we obtain that it must be $0.0283 <n_T < 0.0407$, in order for the $+$ and $\times$ polarisations modes to have the same spectrum. This imposes some restriction on the constant of self-interaction $3.0018\,<\,\xi^2\,< \,3.0025$ of the fermionic source. The most relevant result here obtained is that the intensity of energy density for GW during inflation is $1.25\times 10^{-4} \left(\frac{m}{H}\right)^2 \, < \,\Omega_{GW} \, < \,1.75\times 10^{-4} \left(\frac{m}{H}\right)^2$, where $m$ is the mass of the $1/2$-spin fermionic fields and $H$ the Hubble parameter during inflation. This cut imposes restrictions on the mass of these fields: $\left({m\over H}\right)^2\lesssim  1.1\,h_0^2\times 10^{-11}$.
\end{abstract}
\maketitle

\section{Introduction and Motivation}

Gravitational waves were predicted close to 100 years ago by A. Einstein in the framework of his theory of General Relativity (GR)\cite{einstein}, and revisited by him with N. Rosen twenty years later\cite{einstein1}. In the last few years, Advanced LIGO has opened a new era in the modern astrophysical research\cite{..,..1,..2,..3,..4,..5}, by detecting gravitational waves (GW), coming from collisions of pairs of black-holes. It is expected that in the future, new detectors like LISA, can measure GW emitted during the early universe\cite{haw,haw1,haw2}, in particular, when the universe suffered a quasi-exponential expansion during inflation. During this epoch
the energy density of the Universe was dominated by some scalar field (the inflaton), with negligible kinetic energy density, in
such a way that its corresponding vacuum energy density is responsible for the exponential growth of the scale factor of the
universe. Along this second order phase transition a small and smooth region of the order of size of the Hubble radius, grew so
large that it easily encompassed the comoving volume of the entire presently observed universe, and consequently the observable
universe become so spatially homogeneous and isotropic on scales today of the range [$10^8-10^{10}$] ly. A general prediction of cosmological inflation\cite{infl,infl1,infl2} is the generation of a stochastic background of primordial gravitational waves (GW)\cite{sil,sil1}. Its detection would be of great importance for the understanding and corroborating of inflation during the early phase of the expansion of the universe\cite{mat}. Under the standard model of cosmology plus the theory of inflation, it is very natural to predict the existence of the GW with respect to the background geometry of the universe. In the standard single-field, slow-roll inflationary scenario the tensor fluctuations of the metric are characterized by a nearly scale-invariant power-spectrum on super-Hubble scales. The amplitude of GW signal is described by the tensor-to-scalar ratio $r$, defined as the ratio between the tensor and the scalar power-spectrum amplitudes at a given wave-number $k=k_* \simeq 0.05 \, Mpc^{-1}$, assuming $r=-8 n_T$, where $n_T$ is the tensor spectral index.

However, the existence of a background Riemann geometry should be explained from a more fundamental basis. In a recent work\cite{mb} some of the authors of this work have proposed an unified spinor field (USF) formalism in order to describe non-background relativistic systems which are quantum mechanical in nature. This is a non-perturbative theory of unification for quantised spinor fields on extended manifolds, taking into account the self-interactions of the spinor fields. Each component of spin $\hat{S}_{\mu}$, is defined as the momentum corresponding to the inner dimension $\hat{\Phi}^{\mu}$, such that one can define an universal bi-vectorial invariant: $\left<B\left|\underleftrightarrow{S} \overleftrightarrow{\Phi}\right|B\right> = (2\pi n \hbar) \,\mathbb{I}_{4\times 4}$. The non-perturbative description for GW in this formalism is a nontrivial issue. In this work we develop the dynamics for GW on an extended manifold characterized by the spinor fields
components $\hat{\Psi}^{\mu}$. This theory is described y Sect. II. Firstly we describe the quantum spacetime, in order to introduce the extended theory taking into account the boundary conditions on the minimum action principle. With the aim to illustrate the theory, we describe the dynamics of GW in a model of inflation. This is done in Sect. III. Finally, in Sect. IV we conclude with some final comments.

\section{Quantum structure of spacetime and Unified Spinor Fields}

It is expected that the background spacetime can emerge from the expectation value of a quantum structure of spacetime. In order to propose a
description we shall consider that this spacetime is generated by a base of $4\times 4$-matrices, $\bar{\gamma}^{\alpha}$. The $\bar{\gamma}_{\alpha}= E^{\mu}_{\alpha} \gamma_{\mu}$ matrices which generate the background metric are related by the vielbein $E^{\mu}_{\alpha}$ to basis $\gamma_{\mu}$ in the Minkowsky spacetime (in cartesian coordinates). In this paper we shall consider the Weyl basis, such that:
 $\left\{ \gamma^a,\gamma^b\right\} = 2 \eta^{ab} \mathbb{I}_{4\times 4}$. The elements $\bar{\gamma}^{\mu}$ comply with the Clifford algebra:
\begin{equation}
\bar{\gamma}^{\mu} = \frac{\bf{I}}{3!}\,\epsilon^{\mu}_{\,\,\alpha\beta\nu} \bar{\gamma}^{\alpha}\bar{\gamma}^{\beta}  \bar{\gamma}^{\nu} , \qquad \left\{\bar{\gamma}^{\mu}, \bar{\gamma}^{\nu}\right\} =
2 g^{\mu\nu} \,\mathbb{I}_{4\times 4}, \nonumber
\end{equation}
where ${\bf{I}}={\gamma}^{0}{\gamma}^{1}{\gamma}^{2}{\gamma}^{3}$ is the pseudoscalar, $\mathbb{I}_{4\times 4}$ is the identity matrix, and we define $\noindent{{\epsilon}^{\mu}_{\alpha\beta\nu}=g^{\mu\rho}\epsilon_{\rho\alpha\beta\nu}}$, with:
\begin{equation}
\epsilon_{\rho\alpha\beta\nu}=
\begin{cases}
1 &  \text{if} \ \rho\alpha\beta\nu \ \text{is an even permutation of}\ 0123 \\
-1 & \text{if} \ \rho\alpha\beta\nu \ \text{is an odd permutation of}\ 0123 \\
0 & \text{in any other case},
\end{cases}  \ \ .  \nonumber
\end{equation}
We will consider the Weyl representation:
\begin{eqnarray}
&& \gamma^0= \,\left(\begin{array}{ll}  0 & \mathbb{I} \\
\mathbb{I}  &  0 \ \end{array} \right),\qquad
\gamma^1=  \left(\begin{array}{ll} 0 &  -\sigma^1 \\
\sigma^1 & 0  \end{array} \right),  \nonumber \\
&& \gamma^2= \left(\begin{array}{ll} 0 &  -\sigma^2 \\
\sigma^2 & 0  \end{array} \right),  \qquad \gamma^3= \left(\begin{array}{ll} 0 &  -\sigma^3 \\
\sigma^3 & 0  \end{array} \right),\nonumber
\end{eqnarray}
where the Pauli matrices are
\begin{eqnarray}
&& \sigma^1 = \left(\begin{array}{ll} 0 & 1 \\
1  & 0  \end{array} \right), \quad \sigma^2 = \left(\begin{array}{ll} 0 & -i \\
i  & 0  \end{array} \right), \quad \sigma^3 = \left(\begin{array}{ll} 1 & 0 \\
0  & -1  \end{array} \right). \nonumber
\end{eqnarray}
The idea to introduce these matrices is describe the spinor information in the spacetime and construct a non-commutative basis that can describe quantum effects in a relativistic framework. The Dirac and Majorana matrices are good candidates, but in general it is possible to use any basis that describe a globally hyperbolic spacetime, which is the global geometry necessary to obtain relativistic causality. To describe the quantum structure of space time we shall consider a the variation, $\delta\hat{X}^{\mu}$, of the quantum operator $\hat{X}^{\mu}$:
 \begin{displaymath}
\hat{X}^{\alpha}(x^{\nu}) = \frac{1}{(2\pi)^{2}} \int d^4 k \, \bar{\gamma}^{\alpha} \left[ b_k \, \hat{X}_k(x^{\nu}) + b^{\dagger}_k \, \hat{X}^*_k(x^{\nu})\right],
\end{displaymath}
such that $b^{\dagger}_k$ and $b_k$ are the creation and destruction operators of spacetime, with $\left< B \left| \left[b_k,b^{\dagger}_{k'}\right]\right| B  \right> = \delta^{(4)}(\vec{k}-\vec{k'})$. Moreover, we shall define in the analogous manner the variation $\delta\hat{\Phi}^{\mu}$ of the quantum operator $\hat{\Phi}^{\mu}$ that describes the quantum inner space:
\begin{displaymath}
\hat{\Phi}^{\alpha}(\phi^{\nu}) = \frac{1}{(2\pi)^{2}} \int d^4 s \, \bar{\gamma}^{\alpha} \,\left[ c_s \, \hat{\Phi}_s(\phi^{\nu}) + c^{\dagger}_s \, \hat{\Phi}^*_s(\phi^{\nu})\right],
\end{displaymath}
where $c^{\dagger}_s$ and $c_s$ are the creation and destruction operators of the inner space, such that $\left< B \left| \left[c_s,c^{\dagger}_{s'}\right]\right| B  \right> = \delta^{(4)}(\vec{s}-\vec{s'})$. These operators can be applied to some background  quantum state, and describe a Fock space on an arbitrary Riemann curved space time $\left|B\right>$. The states $\left|B\right>$ do not evolves with time because we shall consider the Heisenberg representation, in which only the operators evolve with time. They comply with
\begin{equation}\label{dif}
\delta\hat{X}^{\mu}\left|B\right> = dx^{\mu}\left|B\right>, \qquad \delta\hat{\Phi}^{\mu}\left|B\right> = d\phi^{\mu}\left|B\right>,
\end{equation}
where $\phi^{\alpha}$ are the four compact dimensions related to their canonical momentum components $s^{\alpha}$ that describe the spin. In order to describe the effective background spacetime, we shall consider the line element
\begin{equation}\label{line}
dS^2 \delta_{BB'}= dx^2 \delta_{BB'} + d\phi^2 \delta_{BB'} = \left<B\right| \hat{\delta X}_{\mu} \hat{\delta X}^{\mu} \left| B'\right> + \left<B\left|\left( \hat{\delta \Phi}_{\mu} \hat{\delta \Phi}_{\nu} \right) \left( \bar{\gamma}^{\mu} \bar{\gamma}^{\nu}\right) \right|B'\right>,
\end{equation}
where
\begin{eqnarray}
\left<B\left|\hat{\delta X}_{\mu} \hat{\delta X}^{\mu} \right|B\right> &=& dx^2\, \,\mathbb{I}_{4\times 4}, \\
\left<B\left|\left( \hat{\delta \Phi}_{\mu} \hat{\delta \Phi}_{\nu} \right) \left( \bar{\gamma}^{\mu} \bar{\gamma}^{\nu}\right) \right|B\right>&=&
\left<B\left|\frac{1}{4} \left\{ \hat{\delta \Phi}_{\mu} ,\hat{\delta \Phi}_{\nu} \right\} \left\{ \bar{\gamma}^{\mu} ,\bar{\gamma}^{\nu}\right\} -\frac{1}{4}
\left[ \hat{\delta \Phi}_{\mu}, \hat{\delta \Phi}_{\nu} \right] \left[ \bar{\gamma}^{\mu}, \bar{\gamma}^{\nu}\right] \right|B\right>\nonumber \\
&=&d\phi^2 \,\mathbb{I}_{4\times 4}. \nonumber
\end{eqnarray}
Therefore, we have obtained an effective line element which comes from the expectation value on the inner product of vectors $\left<B\left|\hat{\delta X}_{\mu} \hat{\delta X}^{\mu} \right|B\right>$, and the inner product of bi-vectors: $\left<B\left|\left( \hat{\delta \Phi}_{\mu} \hat{\delta \Phi}_{\nu} \right) \left( \bar{\gamma}^{\mu} \bar{\gamma}^{\nu}\right) \right|B\right>$.

\subsection{Boundary conditions on Einstein-Hilbert action}\label{ei}

In order to revise the minimum action principle in general relativity, taking into account the boundary conditions produced by quantum effects from a geometrical point of view, we shall consider the Einstein-Hilbert (EH) action for an arbitrary matter lagrangian density ${\cal L}$
\begin{equation}
{\cal I} = \int d^4 x \sqrt{-g} \left[ \frac{R}{2\kappa}+ {\cal L} \right],
\end{equation}
after variation, is given by
\begin{equation}\label{delta}
\delta {\cal I} = \int d^4 x \sqrt{-g} \left[ \delta g^{\alpha\beta} \left( G_{\alpha\beta} + \kappa T_{\alpha\beta}\right)
+ g^{\alpha\beta} \delta R_{\alpha\beta} \right],
\end{equation}
where $\kappa = 8 \pi G$, $R_{\alpha\beta}$ is the background Ricci tensor, $R=g^{\alpha\beta} R_{\alpha\beta}$ is the background scalar curvature, $g_{\alpha\beta}$ is the symmetric background metric tensor, $G$ is the gravitational constant and
\begin{equation}
g^{\alpha\beta} \delta R_{\alpha\beta} =
\left[\delta W^{\alpha}\right]_{||\alpha} - \left(g^{\alpha\epsilon}\right)_{||\epsilon}  \,\delta\Gamma^{\beta}_{\alpha\beta} +
\left(g^{\alpha\beta}\right)_{||\epsilon}  \,\delta\Gamma^{\epsilon}_{\alpha\beta},
\end{equation}
such that $\delta W^{\alpha}=\delta\Gamma^{\epsilon}_{\beta\epsilon}
g^{\beta\alpha}-\delta
\Gamma^{\alpha}_{\beta\gamma} g^{\beta\gamma}$\footnote{We define the covariant derivative of some vector field $\Upsilon^{\beta}$: $\left[\Upsilon^{\beta}\right]_{||\alpha}$
\begin{equation}
\left[\Upsilon^{\beta}\right]_{||\alpha} = \nabla_{\alpha}\Upsilon^{\beta} + \xi^2 \,\delta\Gamma^{\beta}_{\epsilon\alpha}\Upsilon^{\epsilon},
\end{equation}
where $\xi$ is the self-interaction constant, $\nabla_{\alpha}\Upsilon^{\beta}$ is the covariant derivative on the Riemann manifold and $\delta\Gamma^{\beta}_{\epsilon\alpha}$ is the displacement of the manifold with respect to the Riemann one.
In this work we shall extend the Riemann manifold to describe quantum geometric spinor fields
$\hat{\Psi}^{\alpha}$, by using the connections
\begin{equation}\label{ga}
\hat{\Gamma}^{\alpha}_{\beta\gamma} = \left\{ \begin{array}{cc}  \alpha \, \\ \beta \, \gamma  \end{array} \right\}+ \hat{\Psi}^{\alpha}\,g_{\beta\gamma}.
\end{equation}
Here
\begin{equation}\label{uch}
\hat{\delta{\Gamma}}^{\alpha}_{\beta\gamma}=\hat{\Psi}^{\alpha}\,g_{\beta\gamma},
\end{equation}
describes the quantum displacement of the extended manifold with respect to the classical Riemann background, which is described by the Levi-Civita symbols $\left\{ \begin{array}{cc}  \alpha \, \\ \beta \, \gamma  \end{array} \right\}$ in (\ref{ga}). Notice that the background expectation value of the manifold displacement is null: $\left<B\left|\hat{\delta\Gamma}^{\alpha}_{\beta\gamma}\right|B\right>=0$. Therefore, it will be possible to calculate the covariant derivatives of the spacetime operators (\ref{dif}) using the connections (\ref{uch}).}. We shall consider that the flux that crosses the Gaussian hypersurface is due to the existence of some source (quantum in origin), and that it must be described on an extended manifold whose connection is different from the Levi-Civita one. When this flux, that cross the Gaussian-like hypersurface defined on an arbitrary region of the spacetime
is zero, the resulting equations that minimize the EH action are the background Einstein equations: $G_{\alpha\beta} + \kappa\, T_{\alpha\beta}=0$. However, when this flux is nonzero,
one obtains in the last term of the eq. (\ref{delta}), that
$g^{\alpha\beta} \delta R_{\alpha\beta}=\delta\Theta(x^{\alpha})$, such that $\delta\Theta(x^{\alpha})$ is an arbitrary scalar field. This flux becomes zero when there are no sources inside this hypersurface. This arbitrary
hypersurface must be viewed as a 3D Gaussian-like hypersurface situated in any region of
spacetime. Hence, in order to make $\delta {\cal I}=0$ in Equation (\ref{delta}), we must consider the condition: $G_{\alpha\beta} + \kappa T_{\alpha\beta} = \Lambda\, g_{\alpha\beta}$, where $\Lambda$ is the cosmological constant. Additionally, we must require
the constriction $\delta g_{\alpha\beta} \Lambda =
\delta\Theta\, g_{\alpha\beta}$, in order to obtain: $\bar{\delta W}_{\alpha} = \delta W_{\alpha} - \nabla_{\alpha} \delta\Theta$, where the scalar field $\delta\Theta$ complies $g^{\alpha\beta}\nabla_{\alpha}\nabla_{\beta}\delta\Theta\equiv\Box \delta\Theta =0$\cite{rb1,rb2}, in order to the flux of both, $\bar{\delta W}^{\alpha}$ and ${\delta W}^{\alpha}$, to be zero on the Riemann (background) manifold.

Moreover, we can make the transformation
\begin{equation}\label{ein}
\bar{G}_{\alpha\beta} = {G}_{\alpha\beta} - \Lambda\, g_{\alpha\beta},
\end{equation}
and the transformed Einstein equations with the equation of motion for the transformed gravitational waves, hold
\begin{equation}
\bar{G}_{\alpha\beta} = - \kappa\, {T}_{\alpha\beta}. \label{e1} \\
\end{equation}
The Equation (\ref{e1}) provides us with the Einstein equations with cosmological
constant included. Furthermore, since $\delta \Theta(x^{\alpha})\, g_{\alpha\beta} = \Lambda\,\delta
g_{\alpha\beta}$, the existence of the cosmological constant $\Lambda$, is related
to the existence of some source enclosed by $\partial{\cal M}$.

The variations and exact differentials of the operators $\hat{X}^{\mu}$ and $\hat{\Phi}^{\mu}$ on the extended Weylian manifold, are given respectively by
\begin{eqnarray}
\delta\hat{X}^{\mu}\left| B\right> &=& \left(\hat{X}^{\mu}\right)_{\|\alpha} dx^{\alpha}\left| B\right>, \qquad \delta\hat{\Phi}^{\mu} \left| B\right>= \left(\hat{\Phi}^{\mu}\right)_{\|\alpha} d\phi^{\alpha}\left| B\right>, \\
d\hat{X}^{\mu} \left| B\right>&=& \left(\hat{X}^{\mu}\right)_{,\alpha} dx^{\alpha}\left| B\right>, \qquad d\hat{\Phi}^{\mu} \left| B\right>= \left(\hat{\Phi}^{\mu}\right)_{,\alpha} d\phi^{\alpha}\left| B\right>,
\end{eqnarray}
with covariant derivatives
\begin{eqnarray}
\left(\hat{X}^{\mu}\right)_{\|\beta}\left| B\right> &=& \left[\nabla_{\beta} \hat{X}^{\mu} + \hat{\Psi}^{\mu} \hat{X}_{\beta} - \hat{X}^{\mu} \hat{\Psi}_{\beta}\right]\left| B\right>, \\
\left(\hat{\Phi}^{\mu}\right)_{\|\beta}\left| B\right> &=& \left[\nabla_{\beta} \hat{\Phi}^{\mu} + \hat{\Psi}^{\mu} \hat{\Phi}_{\beta} - \hat{\Phi}^{\mu} \hat{\Psi}_{\beta}\right]\left| B\right>.
\end{eqnarray}
In the next subsection we shall use these spinor fields $\hat{\Psi}^{\mu}$ as generators of an extended manifold from which we can develop an
extended theory of general relativity including the quantum effects.

\subsection{Dynamics of spinor field}

The variation of the extended Ricci tensor $\delta{R}^{\alpha}_{\beta\gamma\alpha}=\delta{R}_{\beta\gamma}$: $\delta{R}_{\beta\gamma} = \left(\delta\Gamma^{\alpha}_{\beta\alpha} \right)_{\| \gamma}- \left(\delta\Gamma^{\alpha}_{\beta\gamma} \right)_{\| \alpha}$, is
\begin{eqnarray}
\hat{\delta{R}}_{\beta\gamma} &=& \nabla_{\gamma} \hat{\Psi}_{\beta} - 3 \left(1-\frac{\xi^2}{3}\right) g_{\beta\gamma} \left(\hat{\Psi}^{\nu} \hat{\Psi}_{\nu}\right) \nonumber \\
&-&g_{\beta\gamma} \left(\nabla_{\nu} \hat{\Psi}^{\nu}\right) +\left(1-\frac{\xi^2}{3}\right)\hat{\Psi}_{\beta}\hat{\Psi}_{\gamma}=\hat{U}_{\beta\gamma}+\hat{V}_{\beta\gamma}.
\end{eqnarray}
Notice that, although the background Ricci tensor is symmetric, its variation has both, symmetric and antisymmetric contributions. The former are due to the noncommutative algebra of the operators. The tensors  $\hat{U}_{\beta\gamma}$ and $\hat{V}_{\beta\gamma}$, are the symmetric and antisymmetric parts of $\hat{\delta{R}}_{\beta\gamma}$\cite{mb}:
\begin{eqnarray}
\hat{U}_{\beta\gamma}&=&\frac{1}{2} \left( \nabla_{\beta} \hat{\Psi}_{\gamma}+\nabla_{\gamma} \hat{\Psi}_{\beta}\right)  - g_{\beta\gamma} \left(\nabla_{\nu} \hat{\Psi}^{\nu}\right) \nonumber \\
&-& 3\left(1-\frac{\xi^2}{3}\right) g_{\beta\gamma} \left(\hat{\Psi}_{\nu} \hat{\Psi}^{\nu}\right)  + 3 \left(1-\frac{\xi^2}{3}\right) \left\{\hat{\Psi}_{\beta},\hat{\Psi}_{\gamma}\right\}, \nonumber \\
\hat{V}_{\beta\gamma} &=& -\frac{1}{2} \left( \nabla_{\beta} \hat{\Psi}_{\gamma}-\nabla_{\gamma} \hat{\Psi}_{\beta}\right)+ \frac{3}{2}\left(1-\frac{\xi^2}{3}\right) \left[\hat{\Psi}_{\beta}, \hat{\Psi}_{\gamma}\right] .\nonumber
\end{eqnarray}
The coupling $\xi$ takes into account the self-interaction of the spinor fields $\hat{\Psi}_{\gamma}$ because they alter the spacetime and therefore they alter itself. Its contribution can be seen in those terms which are quadratic in $\hat{\Psi}_{\gamma}$.

The antisymmetric tensor
$\hat{\delta{R}}^{\alpha}_{\alpha\beta\gamma}\equiv \hat{\Sigma}_{\beta\gamma}$, is
\begin{equation}\label{sigma}
\hat{\Sigma}_{\beta\gamma} = \left( \nabla_{\beta} \hat{\Psi}_{\gamma}-\nabla_{\gamma} \hat{\Psi}_{\beta}\right) -\left(1+\xi^2\right) \left[ \hat{\Psi}_{\beta}, \hat{\Psi}_{\gamma} \right].
\end{equation}
Now we can introduce the varied Einstein tensor on the extended manifold by taking into account only the symmetric contributions of the Ricci tensor: $\hat{\delta{G}}_{\beta\gamma}= \hat{U}_{\beta\gamma} - \frac{1}{2} g_{\beta\gamma} \hat{U}$, where $\hat{U}=g^{\alpha\beta} \hat{U}_{\alpha\beta}$:
\begin{eqnarray}
\hat{\delta{G}}_{\beta\gamma}&=&\frac{1}{2} \left( \nabla_{\beta} \hat{\Psi}_{\gamma}+\nabla_{\gamma} \hat{\Psi}_{\beta}\right)+\frac{1}{2} g_{\beta\gamma}\left[ \left(1-\frac{\xi^2}{3}\right)\left(\hat{\Psi}^{\alpha}\hat{\Psi}_{\alpha}\right)\right. \nonumber \\
&+& \left. \left(\nabla_{\nu} \hat{\Psi}^{\nu}\right)\right]
+\frac{1}{2}  \left(1-\frac{\xi^2}{3}\right) \left\{\hat{\Psi}_{\beta}, \hat{\Psi}_{\gamma}\right\} . \label{gg}
\end{eqnarray}
Taking into account the gauge-transformations (\ref{ein}): $\hat{\bar{\delta G}}_{\alpha\beta}= \hat{\delta G}_{\alpha\beta} - g_{\alpha\beta} \hat{\Lambda}$, we obtain that
\begin{equation}\label{la}
\hat{\Lambda} = - \frac{3}{4} \left[ \nabla_{\alpha} \hat{\Psi}^{\alpha} + \left(1- \frac{\xi^2}{3}\right) \hat{\Psi}^{\alpha} \hat{\Psi}_{\alpha}\right].
\end{equation}
Of course, the extended Einstein equations must be evaluated on the background expectation value defined on the Riemann manifold
\begin{equation}
 \left< B\left|\hat{\delta G}_{\alpha\beta} - g_{\alpha\beta} \hat{\Lambda}\right|B \right>=-8\pi\,G
 \left< B\left|\hat{{\delta T}}_{\alpha\beta}\right|B \right>,
\end{equation}
where the variation of the stress tensor $\hat{\delta T}_{\mu\nu}$ with respect to the background, is
\begin{equation}\label{lag}
\delta \hat{T}_{\mu\nu} = 2 \frac{\delta \hat{{\cal L}}}{\delta g^{\mu\nu}} - g_{\mu\nu}\,  \hat{{\cal L}},
\end{equation}
such that the Lagrangian density is given by $ \hat{{\cal L}}= \frac{2}{3\kappa} \hat{\Lambda}$. The expectation value of $\hat{\Lambda}$ provides the contribution of the spinor fields to the cosmological constant, and therefore to the flux that crosses the closed hypersurface $\partial{\cal M}$.
On the other hand, by considering the two antisymmetric tensors $\hat{V}_{\beta\gamma}$ and $\hat{{\Sigma}}_{\beta\gamma}$, we can define other two antisymmetric tensors
$\hat{\cal{N}}_{\beta\gamma} = \frac{1}{2} \hat{V}_{\beta\gamma} - \frac{1}{4}\hat{{\Sigma}}_{\beta\gamma} $ and $\hat{\cal{M}}_{\beta\gamma} = \frac{1}{2} \hat{V}_{\beta\gamma} + \frac{1}{4} \hat{{\Sigma}}_{\beta\gamma} $
\begin{eqnarray}
\hat{\cal{N}}_{\beta\gamma} & = & -\frac{1}{2} \left( \nabla_{\beta} \hat{\Psi}_{\gamma}-\nabla_{\gamma} \hat{\Psi}_{\beta}\right) + \left[ \hat{\Psi}_{\beta}, \hat{\Psi}_{\gamma}\right], \label{n1}\\
\hat{\cal{M}}_{\beta\gamma} & = & \frac{1}{2} \left(1-\xi^2\right) \left[ \hat{\Psi}_{\beta}, \hat{\Psi}_{\gamma}\right], \label{n2}
\end{eqnarray}
such that we can construct the theory with three tensors; the symmetric tensor $\hat{\delta{G}}_{\beta\gamma}$, and the antisymmetric ones $\hat{\cal{N}}_{\beta\gamma}$ and $\hat{\cal{M}}_{\beta\gamma}$. Notice that $\hat{\cal{N}}_{\beta\gamma}$ is free of self-interacctions. Therefore, now we have three tensors which contain all the geometric information of the theory: $\hat{\delta{G}}_{\beta\gamma}$, $\hat{\cal{N}}_{\beta\gamma}$ and $\hat{\cal{M}}_{\beta\gamma}$. We shall require that all these tensors to be conserved on the extended manifold
\begin{equation}\label{din}
\left< B\left|\left(\hat{\bar{\delta G}}^{\beta\gamma}\right)_{\|\gamma}\right|B \right> =0, \quad \left< B\left|\left(\hat{\cal{M}}^{\beta\gamma}\right)_{\|\gamma}\right|B \right>=0, \quad \left< B\left|\left(\hat{\cal{N}}^{\beta\gamma}\right)_{\|\gamma}\right|B \right>=0.
\end{equation}

In the next section we shall see how fermion fields are the source of gravitational waves, and we shall illustrate it with an example.

\section{Gravitational waves from USF}

In order to describe GW, we shall propose the existence of a 2-rank quantum operator $\hat{h}_{\mu\nu}$ such that
\begin{equation}
\frac{1}{2} \Box \hat{\delta h}_{\mu\nu} = -\hat{\delta U}_{\mu\nu}. \label{gw}
\end{equation}
The d'Alambertian in the left side of (\ref{gw}) ensures the wave nature of $\hat{\delta h}_{\mu\nu}$, and the right side is the quantum source of these waves. This source is the variation of the symmetric Ricci tensor
\begin{equation}
 \hat{\delta U}_{\mu\nu}= -\frac{1}{2} \left( \nabla_{\mu} \hat{\Psi}_{\nu} + \nabla_{\nu} \hat{\Psi}_{\mu}\right) + \frac{1}{2} (\xi^2-3) \left\{\hat{\Psi}_{\mu},\hat{\Psi}_{\nu}\right\} + \frac{2}{3} g_{\mu\nu} \hat{\Lambda},
\end{equation}
where, using (\ref{la}), we obtain that $\hat{\Lambda}$ is given by
\begin{equation}
\hat{\Lambda} = - \frac{3}{4} \left[ \nabla_{\alpha} \hat{\Psi}^{\alpha} + \frac{1}{2}\left(1- \frac{\xi^2}{3}\right) g^{\alpha\beta} \left\{\hat{\Psi}_{\alpha}, \hat{\Psi}_{\beta}\right\}\right].
\end{equation}
and takes into account the source of the gravitational waves. The expectation values of $\hat{\delta R}_{\mu\nu}$ and $\hat{\Lambda}$ (on the Riemann background), are
\begin{eqnarray}
\left<B\left| \hat{\delta U}_{\mu\nu}\right|B\right> &=& -\left<B\left| \frac{1}{2} (\xi^2-3) \left\{\hat{\Psi}_{\mu},\hat{\Psi}_{\nu}\right\}\right|B\right> +
\frac{2}{3} g_{\mu\nu} \left<B\left|  \hat{\Lambda} \right|B\right> , \label{eu1}\\
\left<B\left| \hat{\Lambda}\right|B\right>  & = & -\frac{3}{8} \left<B\left|\left(1- \frac{\xi^2}{3}\right) g^{\alpha\beta} \left\{\hat{\Psi}_{\alpha}, \hat{\Psi}_{\beta}\right\} \right|B\right>, \label{eu2}
\end{eqnarray}
where we have used the fact that $\left<B\left|\nabla_{\alpha} \hat{\Psi}^{\alpha} \right|B\right>=0$.
Since it can be proved that $\left<B\left|\hat{\delta U}\right|B\right> = 4 \left<B\left|\hat{\Lambda}\right|B\right>$, the wave equation is
\begin{equation}\label{gw1}
\left<B\left| \Box  \delta \hat{h}_{\mu\nu}\right|B\right> = - 2\kappa \left<B\left| \hat{\delta T}_{\mu\nu}\right|B\right>,
\end{equation}
where $\kappa=8\pi G/c^4$. Therefore, using (\ref{lag}) to obtain the stress tensor in (\ref{gw1}), we obtain
\begin{eqnarray}
\kappa \hat{\delta T}_{\alpha\beta} &=& -\frac{1}{2} \left( \nabla_{\alpha} \hat{\Psi}_{\beta} + \nabla_{\beta} \hat{\Psi}_{\alpha} \right)  - \frac{1}{2} \,g_{\alpha\beta} \,\nabla_{\nu} \hat{\Psi}^{\nu} -\frac{1}{4} g_{\alpha\beta} \left(1-\frac{\xi^2}{3} \right) g^{\mu\nu} \left\{ \hat{\Psi}_{\mu}, \hat{\Psi}_{\nu} \right\} \nonumber \\
&+& \frac{1}{2} \left(\xi^2-3\right) \left\{
\hat{\Psi}_{\alpha}, \hat{\Psi}_{\beta}\right\}.\label{uu}
\end{eqnarray}
We are interested in the study of gravitational waves produced during inflation. In that case we must adopt a co-moving frame with $\hat{U}^{0} \left|B\right> = \mathbb{I}_{4\times4}\left|B\right>$ and $\hat{U}^{j} \left|B\right> = 0$. The TT (Transverse-Traceless) components of these waves: $\hat{\bar{\delta h}}_{ij}=\hat{\delta h}_{ij}-\frac{1}{2} g_{ij}\, \delta h$, have a dynamics governed by the equations
\begin{equation}\label{on}
\Box \left<B\left| \hat{\bar{\delta h}}_{ij}\right|B\right> = -2 \kappa \left<B\left|\hat{S}_{ij}\right|B\right>,
\end{equation}
where $\hat{S}_{\alpha\beta}=\hat{\delta T}_{\alpha\beta}-\frac{1}{2} g_{\alpha\beta}\,\hat{\delta T}$, is
\begin{eqnarray}
-2\,\kappa \hat{S}_{\alpha\beta} &=&  \left( \nabla_{\alpha} \hat{\Psi}_{\beta} + \nabla_{\beta} \hat{\Psi}_{\alpha} \right)  - 2 \,g_{\alpha\beta} \left[\,\nabla_{\nu} \hat{\Psi}^{\nu} + \frac{3}{2}  \left(1-\frac{\xi^2}{3} \right) g^{\mu\nu} \left\{ \hat{\Psi}_{\mu}, \hat{\Psi}_{\nu}\right\} \right] \nonumber \\
&-& \left(\xi^2-3\right) \left\{ \hat{\Psi}_{\alpha}, \hat{\Psi}_{\beta}\right\}.\label{uu}
\end{eqnarray}
Finally, the expectation values of the source terms in the wave equations, are
\begin{equation}
-2\kappa \left<B\left| \hat{S}_{ij} \right|B\right> =  -3 g_{ij}   \left(1-\frac{\xi^2}{3} \right) g^{\mu\nu} \left<B\left| \left\{ \hat{\Psi}_{\mu}, \hat{\Psi}_{\nu} \right\} \right|B\right>-  \left(\xi^2-3\right) \left<B\left|\left\{\hat{\Psi}_{i}, \hat{\Psi}_{j}\right\} \right|B\right>.\label{uuu}
\end{equation}
Notice that the source of $\left<B\left| \hat{S}_{ij} \right|B\right>$ is exclusively given by fermion fields, with mass $m$
\begin{eqnarray}
&& \left< B\left| \left\{\hat{\Psi}_{\mu}({\bf x}, {\bf \phi}), \hat{\Psi}_{\nu}({\bf x}', {\bf \phi}') \right\}\right|B \right> \nonumber \\
&&=  \frac{m^2}{2}\,\,\frac{s^2}{\hbar^2}\, \left\{\bar{\gamma}_{\mu}, \bar{\gamma}_{\nu} \right\}\,\mathbb{I}_{4\times 4} \,  \sqrt{\frac{\eta}{g}}  \,\,\delta^{(4)} \left({\bf x} - {\bf x}'\right) \,\delta^{(4)} \left({\bf \phi} - {\bf \phi}'\right).
\end{eqnarray}
Therefore, the expectation value for the source term in the wave equation (\ref{on}), is
\begin{equation}
-2\kappa \left<B\left| \hat{ S}_{ij} \right|B\right> =  \frac{3m^2 \,\mathbb{I}_{4\times 4} }{2}\,\,\frac{s^2}{\hbar^2}\, \left\{\bar{\gamma}_{i}, \bar{\gamma}_{j} \right\} \, \sqrt{\frac{\eta}{g}} \, \left(\xi^2-3 \right)  \,\,\delta^{(4)} \left({\bf x} - {\bf x}'\right) \,\delta^{(4)} \left({\bf \phi} - {\bf \phi}'\right).\label{ufa}
\end{equation}
The factor $\sqrt{\frac{\eta}{g}}$ takes into account the ratio between the
determinants of tensor metrics in a Minkowsky metric: $\eta$, and in a generic metric: $g$. Furthermore, the factor $\frac{s^2}{\hbar^2}$ takes into account the spin of the fermions in the source.

\section{GW during inflation}

To explore the model we shall consider a de Sitter (inflationary) expansion, where the background spacetime is described by the line element
\begin{equation}\label{m}
dS^2 = a^2(\tau) \left[ d\tau^2 - \delta_{ij} \,dx^i dx^j\right],
\end{equation}
where $\tau$, that runs from $-\infty$ to zero, is the conformal time of the universe, which is considered as spatially flat, isotropic and
homogeneous. If the expansion is governed by the inflaton field $\varphi$, and it is non-minimally coupled to gravity, the universe can be described by the action
\begin{equation}
{\cal I} = \int \,\sqrt{-g}\,\left[\frac{{\cal R}}{2\kappa} + {\cal L}_{\varphi}\right],
\end{equation}
where ${\cal L}_{\varphi}=-\left[{1\over 2} (\varphi')^2 -V(\varphi)\right]$. In a de Sitter expansion the scale factor of the universe is $a(\tau) =-{1\over H\,\tau}$, and the scalar potential is a constant $V(\varphi)= {3 H^2\over \kappa}$. Furthermore, the kinetic component of ${\cal L}_{\varphi}$ is zero, so that $\varphi(\tau)=\varphi_0$.

Now, we must calculate the equation of motion for GW (\ref{on}), during inflation. The $ij$-components of $\left<B\left|\hat{\bar{\delta h}}_{ij}\right|B\right> \equiv \bar{h}_{ij}$, can be expanded as
\begin{small}
\begin{equation}
\bar{h}_{ij} = \frac{1}{(2\pi)^{8}} \int d^4s \int\, d^4k \sum_{n=+,\times}^{} \,^{(n)}\epsilon_{ij} \, \left[\left[C^{(n)}_{k,s}\right]\, \theta_k(\tau)\,e^{i {k}_j{x}^j} \,e^{\frac{{\rm i}}{\hbar} s_{\alpha} \phi^{\alpha}}+ \left[C^{(n)}_{k,s}\right]^{\dagger}\theta^*_k(\tau)\, e^{-i {k}_j{x}^j}\,e^{-\frac{i}{\hbar} s_{\alpha} \phi^{\alpha}} \right],
\end{equation}\end{small}
where $+,\times$ denote the polarisation states in the Transverse-Traceless (TT) gauge, defined by
\begin{equation}
\bar{h}_{0j}=0, \qquad \bar{h}=0.
\end{equation}
For $^{(n)}\epsilon_{00}=0$, the polarization tensor is transverse: $k^{i}\, ^{(n)}\epsilon_{ji}=0$ to the propagation of the wave characterized by the wavenumber components $k^{i}$. Accounting for $h=0$ and $^{(n)}\epsilon_{ij}= \,^{(n)}\epsilon_{ji}$, we can define
\begin{equation}
^{(n)}\epsilon_{11}=\frac{1}{2} \left\{\bar{\gamma}_1, \bar{\gamma}_1\right\}, \qquad  ^{(n)}\epsilon_{22}= - ^{(n)}\epsilon_{11}, \qquad ^{(n)}\epsilon_{12}= \frac{1}{2} \left\{\bar{\gamma}_1, \bar{\gamma}_2\right\},
\qquad  ^{(n)}\epsilon_{21}= \frac{1}{2} \left\{\bar{\gamma}_2, \bar{\gamma}_1\right\}.
\end{equation}
The creation and destruction operators satisfy:
\begin{equation}
\left< B \left|\left[C^{(n)}_{k,s}\right]\left[C^{(n')}_{k',s'}\right]^{\dagger}\right| B \right> = m^2\,\delta_{nn'} \,\delta^{(4)}\left(\vec{k}-\vec{k'}\right) \,\delta^{(4)}\left(\vec{s}-\vec{s'}\right),
\end{equation}
where $m$ is the mass of the fermion fields. By imposing that
\begin{equation}
 \left<B \right|\,\sum_{n=1,2}^{} \,^{(n)}\epsilon_{ij} \,\left[C^{(n)}_{k,s}\right]  =  \left<B \right|\,^{(n)}\epsilon_{ij} \, m\, \theta^*_k(\tau') \,e^{-i k_{j} x^{j'}}\, e^{-\frac{i}{\hbar} \, s_{\alpha}\phi^{\alpha'}},
\end{equation}
we obtain the following differential equation for the time dependent modes, for a de Sitter expansion:\footnote{We use representation on the background spacetime (\ref{m}), of the Dirac functions:
\begin{eqnarray}
\delta^{(4)}\left( x-x'\right) & = & \frac{4\pi }{(2\pi)^4} \int^{\infty}_{-\infty}  \,dk \,\theta_k(\tau)\,\theta^{*}_k(\tau') \int^{\infty}_{-\infty}\,dk\,k^2   \,e^{i \vec{k} . \left(\vec{x}-\vec{x'}\right)}, \nonumber \\
\delta^{(4)}\left(\phi-\phi'\right) & = & \frac{1}{(2\pi)^4} \int^{\infty}_{-\infty}  \,d^{4}s \,   \,e^{i\,{s}_{\alpha}\left({\phi}^{\alpha}-{\phi'}^{\alpha}\right)}.\nonumber
\end{eqnarray}
}
\begin{equation}
\theta{''}_k(\tau) - \frac{2}{\tau}\, \theta'_k(\tau) + k^2 \theta_k(\tau) =- \frac{6 }{\tau^2}\,\frac{s^2}{\hbar^2}\,\left(\xi^2-3 \right)\,\theta_k(\tau) , \label{mod}
\end{equation}
where the Hubble parameter in a de Sitter expansion is a constant $H$, $k^{\mu}=\left(\omega,k^j\right)$ and $x^{\mu}=\left(\tau,x^j\right)$. If we impose the normalization condition for the modes $\theta_k(\tau)=\tau\,\zeta_k(\tau)$:
\begin{equation}
\zeta_k(\tau) \left[{\zeta}^*_k(\tau)\right]' - \zeta^*_k(\tau) \left[{\zeta}_k(\tau)\right]'= i \frac{4\pi}{9},
\end{equation}
we obtain the solution for the Eq. (\ref{mod})
\begin{equation}\label{cc}
\theta_k(\tau)= \frac{i \pi}{3} \, \tau^{3/2} \, {\cal H}^{(2)}_{\nu}[\zeta(\tau)],
\end{equation}
where ${\cal H}^{(2)}_{\nu}[\eta(\tau)]$ is the second kind Hankel function with argument $\eta(\tau)=k\,\tau$, and parameter
\begin{equation}
\nu=\frac{1}{2}\sqrt{9+\left[24 \frac{s^2}{\hbar^2}(3-\xi^2)\right]},
\end{equation}
such that, due to the fact that the tensor index is given by $n_T= 3-2\nu$\cite{mat}, we obtain
\begin{equation}
n_T= 3- \sqrt{9+ 24 \frac{s^2}{\hbar^2} \left(3-\xi^2\right)}.
\end{equation}
Evidence from Planck 2015\cite{planck} estimates the tensor index $n_T = 1- n_s$, in the range
\begin{equation}
0.0283 <n_T < 0.0407,
\end{equation}
for a spectral index: $n_s=0.9655 \pm 0.0062$. For fermions with spin $s=(1/2)\,\hbar$, these values corresponds to
\begin{equation}\label{xx}
3.0018\,<\,\xi^2\,< \,3.0025
\end{equation}
which implies that, if this supposition were correct, $n_T$ would depend very weakly on the coupling $\xi$ of primordial fermion fields, which are the source of GW at cosmological scales. As can be proved, the amplitude of gravitational waves is
\begin{equation}
\Delta_{GW} \simeq \frac{H^2}{2\pi^2 M_p^2},
\end{equation}
which, during inflation would be of the order (for $H\simeq 10^{-5}\,M_p$)
\begin{equation}
\left.\Delta_{GW}\right|_{Infl} \simeq 10^{-11},
\end{equation}
but the present day value should be (for $H_0\simeq 10^{-61}\,M_p$)
\begin{equation}
\left.\Delta_{GW}\right|_{0} \simeq 10^{-123},
\end{equation}
which is the same decay-rate of the cosmological constant\cite{ass}, here the subscript $0$ indicates the today's value. It means that the present day value, $\left.\Delta_{GW}\right|_{0}$, is $10^{-112}$ orders of magnitude smaller than of the value during inflation.

Finally, it is interesting make an estimation of $\Omega_{GW}=\frac{\rho_{GW}}{\rho_c}$, such that $\rho_c=3\,H^2/(8\pi\,G)$ is the critical energy density, and
\begin{equation}
\rho_{GW} = \int d^4x \sqrt{-g}\,\int d^4\phi \,\, \left<B\left|T^0_0\right|B\right> = \frac{5\,m^2}{24\,\kappa}\,\left(\xi^2-3\right).
\end{equation}
Therefore, from (\ref{xx}), we obtain
\begin{equation}
1.25\times 10^{-4} \left(\frac{m}{H}\right)^2 \, < \,\Omega_{GW} \, < \,1.75\times 10^{-4} \left(\frac{m}{H}\right)^2,
\end{equation}
that is consistent with ($h_0$ is the dimensionless Hubble parameter):  $\Omega_{GW} \lesssim  \,1.6\times h_0^2\times 10^{-15}$\cite{prd2006}, for very lightly fermions with mass of the order of $\left({m\over H}\right)^2\lesssim  1.1\,h_0^2\times 10^{-11}$.

\section{Final Comments}

Following the USF theory recently introduced, we have studied the production of GW during inflation. This is an important issue that should be tested in the next years and would give relevant information about the early stages of the inflationary expansion of the universe. One of the important results here obtained is the bound for values for the coupling of the fermion source on cosmological scales. The tensor index in the interval $0.0283 <n_T < 0.0407$, agrees very well with values obtained using slow-roll parameters $\epsilon$ and $\eta$\cite{planck}, and other models as bouncing cosmology\cite{bou}\cite{mab}. This $n_T$-values are the manifestation of the existence of primordial fermionic fields, which should be the source of gravitational waves at cosmological scales. Finally, with respect to the spectrum of GW, it is well known that the extreme of the spectrum corresponding to the very low frequencies $\left(10^{-16} - 10^{-18}\right)\,Hz$, are related to wavelengths with size of the today cosmological sector. They cross the horizon at the beginning of inflation. However, the opposite extreme of the spectrum corresponds to highest frequencies of the unstable modes that cross the horizon at the end of inflation with size of about $10^3$ times (or biggest than) the size of the horizon at this moment, when the universe suffered an expansion close to $e^{N}$ times $1/H$. The minimum wavelength of the spectrum is
\begin{equation}
\lambda_{Inf} \geq  10^{-28} \, e^{N} \ 10^3 \simeq 10^{-25}\times \, e^{N} \, {\rm cm}.
\end{equation}
This means that GW emitted at the end of inflation should have wavelengths larger than $11 \, {\rm cm}$, for $N=60$. This is the more energetic sector of the spectrum and could be detected by LISA (Laser Interferometer Space Antenna) in the future, with the launching by NASA/ESA of three satellites to form an equilateral triangle with a distance of $5 \times 10^6$ kilometers between each of them. Finally, using the cut $\,\Omega_{GW} \lesssim  \,1.6\times h_0^2\times 10^{-15}$ for GW emitted during inflation, we have demonstrated that for $1/2$-spin fermionic fields (which are the source of GW in the USF), the mass of these fields must be very small: $\left({m\over H}\right)^2\lesssim  1.1\,h_0^2\times 10^{-11}$.

\section*{Acknowledgements}

\noindent The authors acknowledge CONICET, Argentina (PIP 11220150100072CO) and UNMdP (EXA852/18), for financial support.

\end{document}